\documentclass[%
 reprint,
 amsmath,amssymb,
 aps,
]{revtex4-2}

\usepackage{graphicx}
\usepackage{dcolumn}
\usepackage{bm}
\usepackage[T1]{fontenc}

\usepackage[colorlinks=true,allcolors=blue]{hyperref}
\hypersetup{breaklinks=true}

\begin{document}

\preprint{APS/123-QED}

\title{Quantum entanglement during single-cycle nonsequential ionization}

\author{Daniel Younis}
\email{dan.younis@outlook.com}
\author{Songbo Xie}
\author{Joseph H. Eberly}
\affiliation{Center for Coherence and Quantum Optics, and Department of Physics and Astronomy,\\University of Rochester, Rochester, New York 14627, USA}

\date{14 March 2024}

\begin{abstract}
In order to elucidate the correlated motion of atomic electrons, we investigate the attosecond-scale dynamics of their entanglement arising due to nonsequential ionization driven by a strong, linearly-polarized laser field. The calculation is based on numerical integration of the time-dependent Schr\"{o}dinger equation for helium irradiated by a one-cycle, near-infrared field whose intensity is in the neighborhood of $1\textrm{ PW/cm}^2$. The entanglement measure (Schmidt weight) is resolved on a sub-cycle timescale, and its key dependency on the field profile is exposed for the first time by tuning the carrier-envelope phase (CEP) to control the ionization-recollision timing. We find that between CEP cases, this can result in a $20\%$ enhancement in the peak entanglement. A connection is made between the entanglement, the probability current, and the correlation coefficient for the two electron momenta, providing new insights into the nonsequential ionization mechanism.
\end{abstract}

\maketitle


\section{Introduction}
Electron correlations are ubiquitous in nature, and are responsible for a variety of material effects including magnetism \cite{Kakehashi:2006}, superconductivity \cite{Yanase:2003}, and heavy-fermion behavior \cite{Stewart:1984}. In atomic vapors irradiated by laser light, a fundamental process showing correlated electron emission is nonsequential double ionization (NSDI) \cite{LHuillier:1983, Fittinghoff:1992, Kondo:1993, Walker:1994, Talebpour:1997, Weber:2000, Becker:2005, Becker:2008, Corkum:2011}.

\begin{figure}[b]
    \centering
    \includegraphics[width=0.425\textwidth]{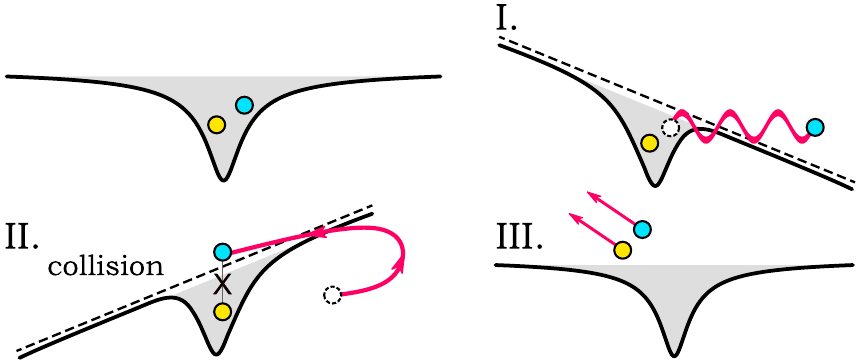}
    \caption{Illustration of the RII (recollision-impact ionization) channel of nonsequential double ionization. The process begins with (I) tunneling emission of one electron through the atom\,$+$\,field potential, to (II) its field-guided return and subsequent collision with the ion in the following half-cycle, and finally to (III) second ionization.}
    \label{fig:Diagram-NSDI}
\end{figure}

Many double-ionization channels are classified as nonsequential \cite{Becker:2005}, but the predominant mechanism is the so-called three-step process \cite{Becker:2012} in which (i) a first electron tunnels through the field-depressed atomic potential, (ii) it is then driven back to the parent ion once the oscillating field reverses sign, and (iii) it scatters inelastically off of the parent ion, transferring energy which collisionally detaches or excites a second electron. The former case is called recollision-impact ionization (RII), and it is illustrated in Fig.~\ref{fig:Diagram-NSDI}. In the latter case, the second electron is more readily field-ionized from its excited state, and the sub-channel is therefore termed recollision-excitation with subsequent ionization (RESI) \cite{Feuerstein:2001, Eremina:2003, Shaaran:2010, Mauger:2012, Yang:2021}. The RII and RESI channels are respectively referred to as the direct and delayed pathways of double ionization.

The \emph{nonsequential} nature of this interaction is contrasted by the sequential picture in which the inter-electron ($e$-$e$) potential is assumed to be negligible, relative to the field interaction potential, so the electrons are detached independently and in-sequence with the external radiation field attaining its maximum strength. The experimental signature of NSDI is a doubly-charged ion yield that exceeds the value calculated under the assumption of sequential ionization, based on the Ammosov--Delone--Krainov (ADK) rate formula \cite{Ammosov:1986}. Early experimental studies of NSDI in helium \cite{Fittinghoff:1992, Kondo:1993, Walker:1994} revealed in some cases a \emph{six order of magnitude enhancement} in the $\textrm{He}^{2+}$ yield. The NSDI signature has been observed in noble gas atoms from helium to xenon \cite{LHuillier:1983, Fittinghoff:1992, Kondo:1993, Walker:1994, Talebpour:1997}.

A consequence of correlated emission is that the electrons become entangled, to a greater degree than in the initial quantum state, as demonstrated by studies of continuous \cite{Grobe:1994, Liu:1999, Omiste:2019, Christov:2019} as well as discrete \cite{Maxwell:2022} dynamical variables. Through numerical calculations, it has also been determined that classical correlations are sufficient to account for the main experimental features, in the ion yield and photoelectron momentum distribution data, when the intensity exceeds $\sim\! 10^{14}\textrm{ W/cm}^2$ \cite{Ho:2005}.

The nature of NSDI under a single optical cycle is questionable in regard to the multi-cycle limit \cite{Bergues:2012, Bergues:2015, Chen:2017, Liu:2021}. The single-cycle regime is particularly important for direct \emph{quantitative} confirmation of experiment, where results are presently limited due to the computational burden of modeling multiple-electron dynamics. This limitation becomes worse in full $3n$-dimensionality, for $n$ electrons, and under irradiation by intense, multi-cycle fields of near-infrared wavelength \cite{Smyth:1998, Bauer:2017}. The one-cycle regime is the cleanest to examine from a theoretical perspective as in that case the driving laser field allows for only one or two principal $e$-$e$ collision events. Kinematically complete experiments with ultrashort (near-single-cycle) laser pulses, with full control of the field, have long been challenging, but are gradually becoming accessible \cite{Bergues:2012, Bergues:2015, Camus:2012}. As a step toward controlling the correlated electron motion, it is necessary to examine the buildup of entanglement on the sub-cycle (attosecond) timescale, which has not been considered in detail before.

The interesting open questions about NSDI in the one-cycle regime include the following \cite{Bergues:2012, Bergues:2015, Chen:2017, Liu:2021}. First, one can ask whether strong correlations even appear in this domain, or if the possibility of recollision is critically dependent on multiple cycles. Second, on a sub-cycle timescale, to what degree is the $e$-$e$ entanglement sensitive to variations in the field profile? For instance, will the end-of-pulse entanglement measure be maximized if the field induces a single high-momentum $e$-$e$ collision, or multiple collisions of comparatively low momentum? Also, can we predict when the entanglement will increase based on the field profile and the simple qualitative picture of electron acceleration and recollision?

We can provide answers to all of these questions. We can increase understanding by analyzing the time development of an entanglement measure (Schmidt weight) as the ionization-recollision dynamics is altered. By tuning the carrier-envelope phase of the single-cycle pulse, we can vary the momentum gain of the first liberated electron before it experiences a recollision event with a second, still bound electron. We find that in the early stages of ionization, the entanglement increases in a predictable way with the laser field maxima. An analysis of probability current at different locations enables a connection to be established with the wavefunction components in space.

The remainder of this article is organized as follows. In Sec.~\ref{sec:Methods}, we discuss the atomic model employed as well as the numerical methods for time-propagating the wavefunction and calculating the entanglement. In Sec.~\ref{sec:Results}, the calculation results are presented, and the implications for entanglement during nonsequential ionization are analyzed in depth. Finally, in Sec.~\ref{sec:Conclusion}, we summarize and conclude this work. Except where otherwise indicated, atomic units (a.u.)~are employed.

\section{Numerical Methods\label{sec:Methods}}
\subsection{Atomic and field potentials}
The model Hamiltonian describes two electrons each constrained to move in one dimension $\vec{x}=(x_1,x_2)$ which is aligned with the linear field polarization axis,
\begin{align}
    \hat{H}&(\vec{x},\vec{p},t) = \sum_{n=1}^{2}\bigg[\frac{1}{2}\hat{p}_n^2 - ZV(x_n)\bigg] \label{eqn:AE-Hamiltonian} \\
    &+ V(x_1-x_2) + (x_1+x_2)E(t),\quad \hat{p}_n=-i\frac{\partial}{\partial x_n}. \nonumber
\end{align}
Systems of aligned electrons have been the subject of numerous investigations \cite{Haan:2002, Lein:2000, Becker:2012} as they reduce the computational demand of the $e$-$e$ interaction while producing sensible results that agree qualitatively with experiment, particularly of photoelectron momentum distributions \cite{Lein:2000, Younis:2023}. Other, more complex reduced-dimensionality models exist, such as that of Ruiz \emph{et al.}~\cite{Ruiz:2006} which treats the $2e$ center-of-mass motion as polarization-aligned, and the Eckhardt--Sacha model \cite{Sacha:2001} which accounts for transverse electron motion to a degree.

\begin{figure}[t]
    \centering
    \includegraphics[width=0.3\textwidth]{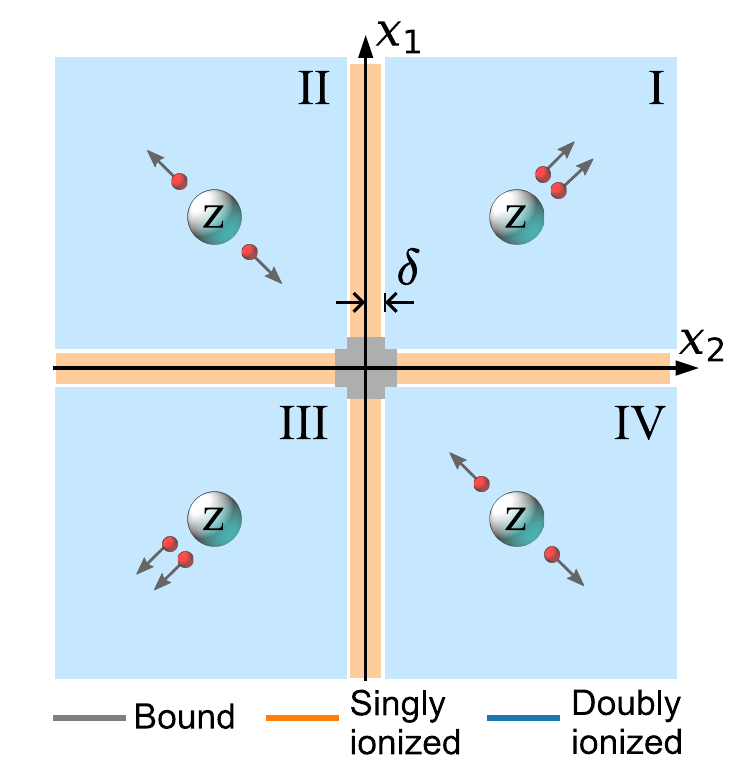}
    \caption{Diagram of the aligned $2e$ position space, illustrating the regions where the wavefunction is bound (gray), singly ionized (orange), and doubly ionized (blue). In Quadrants I and III $(x_1 x_2>0)$, the electrons are co-located on the same side of the nucleus, and in Quadrants II and IV $(x_1 x_2<0)$ they are on opposite sides. The threshold for ionization is characterized by the distance $\delta$ away from the nucleus.}
    \label{fig:Space-2e}
\end{figure}

In Fig.~\ref{fig:Space-2e}, an illustration of the $2e$ position-space is provided, along with the interpretation of each region. In Eq.~(\ref{eqn:AE-Hamiltonian}), $Z=+2$ is the nuclear charge, $V(x)=(x^2+a^2)^{-1/2}$ is the softened Coulomb potential \cite{Javanainen:1988, Su:1991} with core radius $a=1\textrm{ a.u.}$, and $E(t)$ is the laser electric field in dipole approximation. These parameter values correspond to a helium atom with a ground-state energy of $-2.238\textrm{ a.u.}$~($\approx -60.91\textrm{ eV}$).

The laser electric field is derived from a vector potential with a sine-squared envelope:
\begin{align}
    &A(t)=-(E_0/\omega)\sin^2(\pi t/T_p)\sin(\omega t + \phi), \label{eqn:field} \\[0.25em]
    &E(t)=-dA(t)/dt, \nonumber
\end{align}
of amplitude $E_0$, frequency $\omega$, pulse duration $T_p$, and carrier-envelope phase (CEP) $\phi$. The negative sign for $A(t)$ is a matter of convention, and for $\phi=0$ the field initially accelerates the electrons to the left of the nucleus $(x_{1,2}<0)$. Below, we will consider pulse durations $T_p=2\pi N/\omega$ of $N=1$ and $3$ cycles, and the behavioral transition between the single- and multi-cycle regimes.

The Hamiltonian (\ref{eqn:AE-Hamiltonian}) is symmetric under electron exchange and, therefore, with the laser magnetic field being neglected (dipole approximation), the interaction engages only the subset of states with opposite spin projections. Further, note that in the absence of the $e$-$e$ interaction term $V(x_1-x_2)$, the Hamiltonian factorizes into one-electron Hilbert spaces $\hat{H}=\hat{H}_1\otimes\hat{H}_2$. Thus, the $2e$ degree of entanglement is closely related to the overall magnitude of this term.

\subsection{Wavefunction propagation}
The time-dependent Schr\"{o}dinger equation (TDSE) $i\partial\Psi/\partial t=\hat{H}\Psi$ is integrated numerically for the two-electron wavefunction $\Psi(\vec{x},t)$. This is accomplished using the Peaceman--Rachford alternating direction method \cite{Peaceman:1955, Bauer:2017} which weaves the two spatial dimensions in Crank--Nicolson fashion every time-step $\delta t$ according to the scheme:
\begin{align}
    &\Psi\big(\vec{x},t+\tfrac{1}{2}\delta t\big)\gets \big(1+\tfrac{i}{2}\hat{H}_2\delta t\big)^{-1} \big(1-\tfrac{i}{2}\hat{H}_1\delta t\big)\Psi(\vec{x},t), \label{eqn:PRADI-method}
    \\
    &\Psi(\vec{x},t+\delta t)\gets \big(1+\tfrac{i}{2}\hat{H}_1\delta t\big)^{-1} \big(1-\tfrac{i}{2}\hat{H}_2\delta t\big)\Psi\big(\vec{x},t+\tfrac{1}{2}\delta t\big), \nonumber
\end{align}
where $\hat{H}=\hat{H}_1+\hat{H}_2$, and the $\hat{H}_n=-\tfrac{1}{2}[\partial^2/\partial x_n^2-V(\vec{x},t)]$ are symmetric sub-Hamiltonians with total potential energy $V(\vec{x},t)$. Scheme (\ref{eqn:PRADI-method}) is implicit and unconditionally stable with a single-step error of $\mathcal{O}(\delta t^3)$. In the position basis, the discrete differential operators for $\hat{H}_1$ and $\hat{H}_2$ are represented by tridiagonal matrices which are sparse and therefore amenable to efficient numerical manipulation. However, while convenient, operator-splitting does incur an error cost on unitarity that increases with the degree of non-commutativity, $[\hat{H}_1,\hat{H}_2]\neq 0$. For the aligned two-electron Hamiltonian, $[\hat{H}_1,\hat{H}_2]$ does not vanish identically, but convergence tests indicate that unitarity is nevertheless well preserved.

\subsection{Degree of entanglement}
To compute the degree of $e$-$e$ entanglement, use is made of the measure defined by Grobe \emph{et al.}~\cite{Grobe:1994}. To summarize its derivation, the $2e$ wavefunction can be expressed as a sum over a single index \cite{Everett:1957}:
\begin{equation}
\Psi(\vec{x},t)=\sum_{n}C_n(t)\,\sigma_n(\vec{x})
\label{eqn:canonical-wavefunction}
\end{equation}
where, for integer $n$,
\begin{equation}
\sigma_n(\vec{x})=\frac{1}{\sqrt{2}}
\begin{vmatrix}
\psi_{n\alpha}(x_1) & \psi_{n\beta}(x_1) \\[0.35em]
\psi_{n\alpha}(x_2) & \psi_{n\beta}(x_2)
\end{vmatrix}
\end{equation}
are Slater determinants of one-electron orbitals $\psi_{n\alpha}(x_j)\equiv\psi_n(x_j)\,\alpha(j)$ ($\alpha,\,\beta$ are the spin components), and $C_n(t)$ are time-dependent coefficients. The entanglement measure is, loosely speaking, associated with the total number of determinants $\sigma_n(\vec{x})$ required to construct the full $2e$ wavefunction. Along this line, note the squared coefficients in Eq.~(\ref{eqn:canonical-wavefunction}) are probability weights that are suitably normalized, $\sum_n|C_n|^2=1$. The \emph{average} probability is then $\sum_n|C_n|^4$ and its inverse provides a measure of the \emph{number} of non-zero probabilities. This leads to the so-called Schmidt weight
\begin{equation}
    K\equiv\bigg[\sum_n|C_n|^4\bigg]^{-1}\label{eqn:Schmidt-weight}
\end{equation}
as an entanglement measure. In practice, this quantity is computed through the single-particle density operator
\begin{equation}
    \rho(x,x';t)=\int dx_2\,\Psi(x,x_2;t)\Psi^*(x',x_2;t) \label{eqn:density-operator}
\end{equation}
wherein the coordinate degree of freedom (and spin, if relevant) of one electron has been integrated out (summed over). The eigenvalues of this Hermitian operator $\rho(x,x';t)$ are the desired probability weights $|C_n|^2$. Note that while the matrix representation of the density operator contains $n_x$ eigenvalues, where $n_x\sim\mathcal{O}(10^3)$ is the number of discrete spatial points, the vast majority of them are negligible. For the parameters under consideration, convergence is obtained by calculating only the $20$ largest eigenvalues.

In what follows, we will use the normalized entanglement monotone for $M$-level systems \cite{Qian:2018}:
\begin{equation}
    Y_M(K)\equiv 1-\sqrt{\frac{M-K}{K(M-1)}},\quad 0\leq Y\leq 1, \label{eqn:normalized-entanglement}
\end{equation}
which is given approximately as $Y\approx 1-\sqrt{1/K}$ for an atomic model with $M\gg 2$ levels, as considered here. The entanglement measure $Y_M(K)$ is implicitly time-dependent as it is derived from $\rho(x,x';t)$, and will henceforth be denoted $Y(t)$. Last, a meaningful comparison between entanglement values at different times can only be made if the wavefunction remains normalized. Thus, while we do employ absorbing boundary conditions for $\Psi(\vec{x},t)$, the numerical domain is sufficiently large that the loss of probability density is negligible. Note that in the multi-cycle regime, the pioneering study by Liu \emph{et al.}~\cite{Liu:1999} is based on a spatial two-zone method \cite{Grobe:1999} to maintain a feasible numerical volume and efficiently propagate the wavefunction components in the far field.

\section{Results\label{sec:Results}}
\subsection{Signature of NSDI in single-cycle ionization}
Historically, the first evidence that $e$-$e$ correlations play a significant role during ionization appeared in the intensity-dependent yield of doubly-charged ions \cite{Walker:1994}. The characteristic shape of the logarithmic curve is a ``knee'' that signifies a markedly larger yield than normally anticipated at moderate intensities, which for helium is within the range $0.5$\,--\,$1\textrm{ PW/cm}^2$.
\begin{figure}[t]
    \centering
    \includegraphics[width=0.475\textwidth]{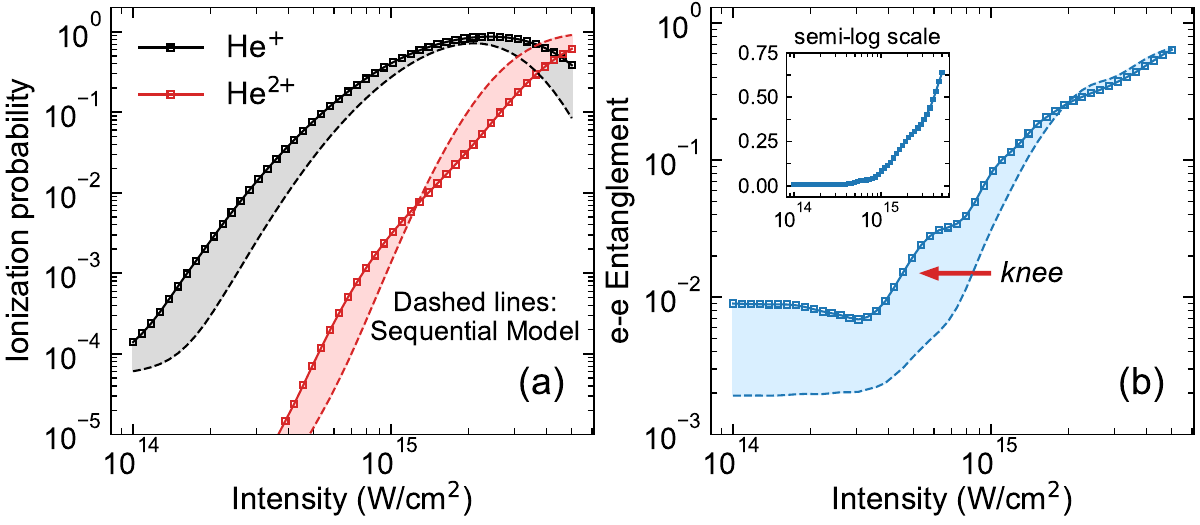}
    \caption{Results for model helium irradiated by a single-cycle $780\textrm{ nm}$-wavelength field. (a) Singly- and doubly-ionized yields vs.~laser intensity. (b) Two-electron entanglement vs.~laser intensity (inset: semi-log scale). Dashed curves in (a), (b): Ionization probability\,/\,$e$-$e$ entanglement computed for a model atom exhibiting more sequential-like ionization behavior.}
    \label{fig:NSDI-signature}
\end{figure}

Inter-electron correlations are significant in ionization with a single-cycle field as well. To demonstrate this, we have performed TDSE integrations of our model helium atom irradiated by a one-cycle $780\textrm{ nm}$-wavelength field for 50 values of intensity ranging from $0.1$\,--\,$5\textrm{ PW/cm}^2$. The single- and double-ionization probabilities are presented in Fig.~\ref{fig:NSDI-signature}{\color{blue}(a)}. The criteria for computing the $\textrm{He}^{+}$/$\textrm{He}^{2+}$ population fractions from the wavefunction are based on the orange/blue regions of the domain in Fig.~\ref{fig:Space-2e}, with cutoffs of $\delta=10\textrm{ a.u.}$~away from the nucleus,
\begin{align*}
    &\textrm{Single ionization (SI):}\,\,|x_i|\geq\delta\,\,\textrm{and}\,\,|x_j|<\delta,
    \\
    &\textrm{Double ionization (DI):}\,\,|x_{i,j}|\geq\delta.
\end{align*}
To highlight the ionization enhancement due to $e$-$e$ interaction, we also considered a model atom which exhibits stronger sequential behavior. It was created by artificially reducing the amplitude of $V(x_1-x_2)$ in Eq.~(\ref{eqn:AE-Hamiltonian}) by $50\%$ while also increasing the core radius to $a=1.184\textrm{ a.u.}$~to maintain the same ground-state energy. The SI/DI probabilities of this more ``Sequential Model'' are overlaid in Fig.~\ref{fig:NSDI-signature}{\color{blue}(a)} (dashed lines) to accentuate the knee structure. Below $\approx 1\textrm{ PW/cm}^2$, $e$-$e$ interaction enhances the $\textrm{He}^{2+}$ population fraction. For higher intensities, the Sequential Model has a higher DI yield, which is expected as the models have different thresholds for second ionization. The ratio of DI probabilities between the two models is $6.2$ at an intensity of $0.5\textrm{ PW/cm}^2$.

In Fig.~\ref{fig:NSDI-signature}{\color{blue}(b)}, the end-of-pulse normalized entanglement $Y(T_p)$ vs.~intensity is presented, and one observes a similar knee structure in the sub-$1\textrm{ PW/cm}^2$ intensity region. By contrast, the Sequential Model (dashed line) does not display the characteristic knee shape around $0.5\textrm{ PW/cm}^2$. The ratio of entanglement values between the two models is $4.7$ at the starting intensity of $0.1\textrm{ PW/cm}^2$. This data is perhaps the strongest indicator that multiple-ionization enhancement is due to $e$-$e$ interaction and entanglement \cite{Grobe:1994, Liu:1999}, and forms the basis of our subsequent analysis into electron correlations on short timescales.

\subsection{Sub-cycle entanglement dynamics}
The time development of nonsequential ionization is not easily interpreted from the dynamical behavior of the wavefunction alone. With every field half-cycle, jets of probability current are released into Quadrants I \& III of Fig.~\ref{fig:Space-2e} \cite{Haan:2002} that are known to arise from both sequential and nonsequential double-ionization pathways \cite{Younis:2023}. However, the relative contributions are not easily determined without resorting to quantum trajectory methods. Indeed, such classical \cite{Wang:2010} and quasi-classical \cite{Chen:2017} approaches have found remarkable success not only in reproducing experimental results, but in providing intuitive pictures of the $2e$ interaction \footnote{Quasi-classical models typically have \emph{ad hoc} tunneling assumptions that are built into the time evolution stochastically.}.

Given a classically-computed ensemble of double ionization events, the associated sequential and nonsequential $2e$ trajectories, which are oftentimes statistically weighted, can be distinguished and unambiguously compared quantitatively. By contrast, we show in this subsection that the entanglement---here a purely quantum-mechanical measure---when analyzed temporally provides valuable information on the timing of $e$-$e$ collisions. It is useful to consider the buildup of entanglement within the first optical cycle, an aspect which has received attention in the context of electron-ion entanglement \cite{Czirjak:2013, Majorosi:2017}.

\subsubsection{Relation to the field profile}
\begin{figure}[t]
    \centering
    \includegraphics[width=0.475\textwidth]{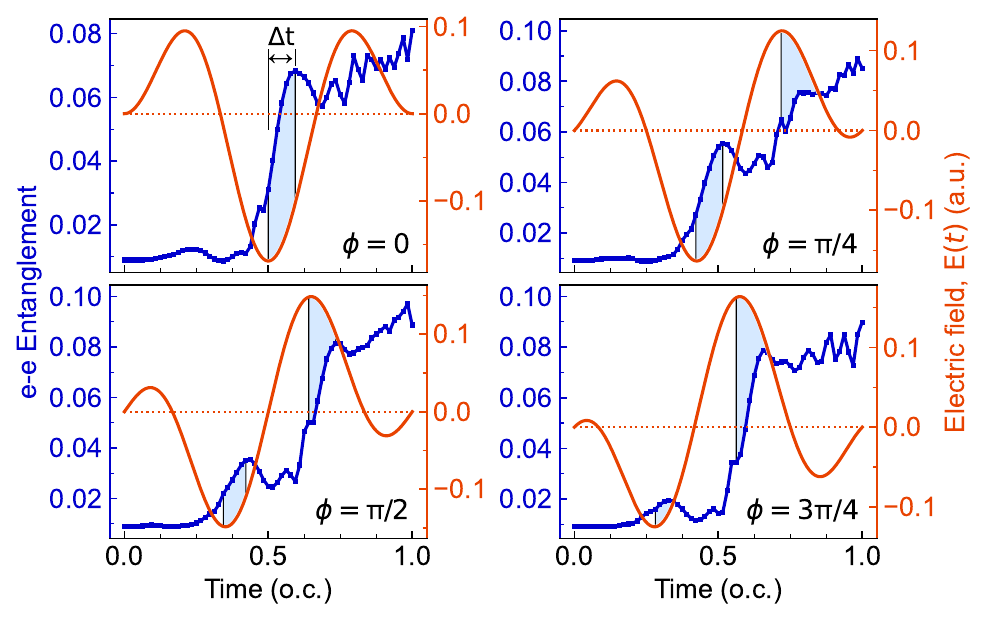}
    \caption{Time development of $e$-$e$ entanglement (blue) at four different CEP values $\phi$, as labeled, and the corresponding laser electric field profile (orange). The orange dotted line marks $E(t)=0$.}
    \label{fig:Schmidt-CEP-r3}
\end{figure}

Figure \ref{fig:Schmidt-CEP-r3} shows the normalized entanglement as a function of time $Y(t)$ for different CEP values $\phi$ in multiples of $\pi/4$. Overlaid in each subplot is the corresponding laser electric field profile $E(t)$, which is only active for one optical cycle (henceforth abbreviated ``o.c.''). Here, the wavelength is $780\textrm{ nm}$, and the intensity is $1\textrm{ PW/cm}^2$. The entanglement of the initial (ground) state is $Y(0)\approx 9.01\times 10^{-3}$. Note the entanglement time evolution is invariant for $\phi\to\phi+\pi$, which simply reverses the sign of $E(t)$.
\begin{figure}[t]
    \centering
    \includegraphics[width=0.425\textwidth]{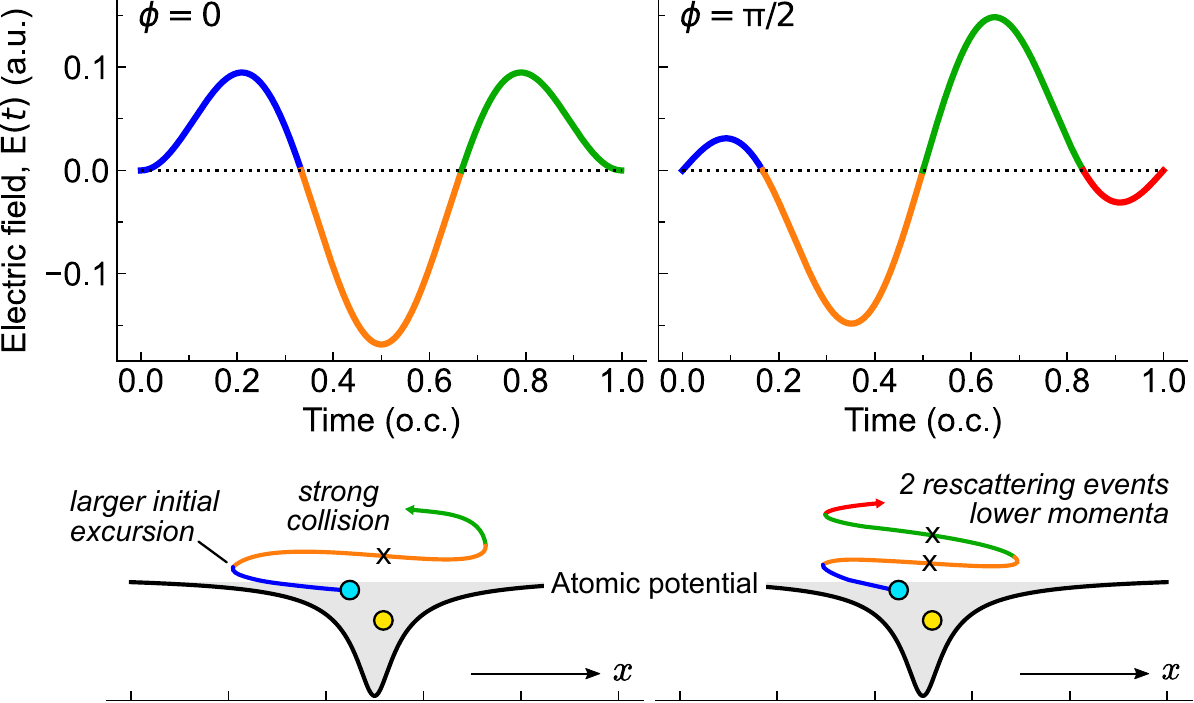}
    \caption{Top panels: Color coding the different sign phases of the laser electric field, $E(t)$, of two CEP values $(\phi=0,\,\pi/2)$. Bottom panels: Diagram of classical electron trajectories, illustrating how the acceleration and collision (\textsf{x}) dynamics vary during each of the phases.}
    \label{fig:Diagram_e-e_collision-r3}
\end{figure}

In the $\phi=0$ case, observe that there is a steep increase in $Y(t)$ shortly after the field attains its maximum strength. This behavior is readily understood qualitatively from the field profile and its influence on the $2e$ dynamics. The first crest in $E(t)$, occurring within $t\approx 0.34\textrm{ o.c.}$, causes the $2e$ probability density to eject predominantly into the $x_i\ll 0$, $x_j\approx 0$ regions of space, which signify single ionization. As $E(t)$ reverses sign over the period of $t\approx 0.34$\,--\,$0.67\textrm{ o.c.}$, the detached electron is then accelerated back toward the nucleus where an $e$-$e$ collision event takes place and, consequently, the entanglement rises sharply.

In the $\phi=\pi/4$ and $\pi/2$ cases, we observe two moments in time at which $Y(t)$ increases in a step-like fashion, and they both follow shortly after the field attains its maximum amplitude. Once more, this behavior is explained by the sign reversal of $E(t)$ leading to $(e,2e)$ rescattering and ionization. Here, however, it becomes apparent that the entanglement increases are regular in the sense that they seem to always occur approximately $\Delta t=1/10$th of a cycle after a maximum in field strength. In Fig.~\ref{fig:Schmidt-CEP-r3}, $\Delta t$ is the field-to-entanglement peak-to-peak time interval shaded in blue.

Note the absolute field strength for $\phi=\pi/2$ has two maxima located symmetrically about $t=0.5\textrm{ o.c.}$, whereas for $\phi=0$ there is a single global maximum and, consequently, a principal $e$-$e$ collision event shortly after the first half-cycle. However, the peak entanglement in the $\phi=\pi/2$ case is $20\%$ greater than in the $\phi=0$ case, suggesting that multiple field reversals may enhance the entanglement, even if the momentum prior to collision is reduced as a result. To illustrate this point, in Fig.~\ref{fig:Diagram_e-e_collision-r3} we show diagrammatically how the $\phi=0$ and $\pi/2$ field profiles influence the acceleration and recollision dynamics of a classical electron pair. In the $\phi=0$ case, the first electron recollides with comparatively higher momentum due to the longer initial ionization-acceleration phase (highlighted in blue). However, in the $\phi=\pi/2$ case, due to the number of field sign reversals, it is more likely that there will be two rescattering events. This simple qualitative picture is in agreement with the observed step-like entanglement increases in Fig.~\ref{fig:Schmidt-CEP-r3}.

\begin{figure}[t]
    \centering
    \includegraphics[width=0.425\textwidth]{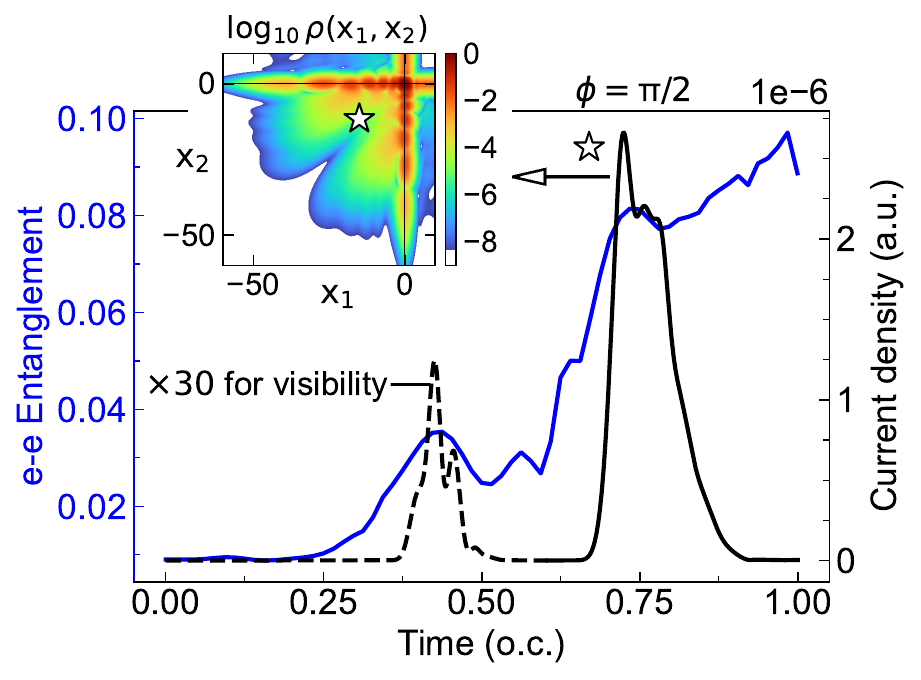}
    \caption{Time development of $e$-$e$ entanglement (blue) and the current density magnitude $|\vec{\jmath}(t)|$ recorded at the points $\vec{x}_d=(\pm 15,\pm 11.6)\textrm{ a.u.}$~(black, $+$/$-$ for dashed/solid). The inset shows the probability density $\log_{10}\rho$ at the time instant when the current density signal (solid line) has attained its maximum value, and the star indicates the corresponding position at which it was recorded.}
    \label{fig:J-Schmidt-DI-Signals}
\end{figure}

To better understand the origin of the entanglement's sensitivity to the field profile, the probability current density $\vec{\jmath}(t)=\tfrac{-i}{2}(\Psi^*\nabla\Psi - \Psi\nabla\Psi^*)|_{\vec{x}_d}$ was recorded at different points $\vec{x}_d$ surrounding the nucleus. It was discovered that in specific regions of space, the $|\vec{\jmath}(t)|$ signals increase in exact coincidence with the entanglement. Figure \ref{fig:J-Schmidt-DI-Signals} shows $Y(t)$ from the $\phi=\pi/2$ case overlaid with the current density magnitude recorded at the points $\vec{x}_d=(\pm 15,\pm 11.6)\textrm{ a.u.}$ Notice how the maxima between the two signals $Y(t)$ and $|\vec{\jmath}(t)|$ occur at the same instants of time. These detection points are significant as they are in the path of double-ionization jets \cite{Haan:2002, Younis:2023}, which are ejected close to the $x_1=x_2$ diagonal with every passing half-cycle (see the inset of Fig.~\ref{fig:J-Schmidt-DI-Signals}). As a result, we conclude that nonsequential ionization produces double-ionization jets of strongly entangled electrons, with this particular case showing three-fold increases (approximately) with the emission of each jet. This behavior only occurs in the single- (or near-single) cycle regime, as previous studies \cite{Grobe:1994, Liu:1999} indicate a saturation in the entanglement measure over time, as the bound population is depleted. As a final remark, we note that increases in the entanglement measure are in fact \emph{a consequence} of increased probability amplitude near to the $x_1=x_2$ diagonal, as it maximizes the only non-factorizable term in the Hamiltonian, $V(x_1-x_2)$.

\subsubsection{Relation to the momentum correlation coefficient}
Our calculations also reveal that the Schmidt weight measure is closely related to a well-known statistical measure, the Pearson correlation coefficient (PCC) \cite{Pearson:1895}. Since entanglement is a type of correlation, this similarity can be expected. From a qualitative standpoint, the PCC between the positions or momenta of the electrons offers a clearer picture of the ionization dynamics. Recently, it has been proposed to experimentally characterize entanglement in qubit systems \cite{Maccone:2015}, which has stimulated extensions to bipartite qutrits \cite{Ghosh:2018} and, incrementally, higher-dimensional states \cite{Jebarathinam:2020} (a comprehensive review is also contained in Ref.~\cite{Jebarathinam:2020}). To the best of our knowledge, until now the PCC has not been studied in connection with atomic electron entanglement, although it has recently been employed in the analysis of molecular double ionization \cite{Zhao:2018}. Accordingly, we consider the correlation coefficient between the two electron momenta $(p_1,p_2)$ given by the formula
\begin{equation}
    \mathcal{C}(p_1,p_2) = \frac{\textrm{cov}(p_1,p_2)}{\sigma(p_1)\,\sigma(p_2)}.
\end{equation}
The covariance and standard deviations are given, respectively, by the standard formulae:
\begin{align*}
    &\textrm{cov}(p_1,p_2) = \langle p_1 p_2\rangle - \langle p_1\rangle\langle p_2\rangle, \\[0.25em]
    &\sigma(p_n)=\sqrt{\langle p_n^2\rangle - \langle p_n\rangle^2},
\end{align*}
and $\langle f(\vec{p})\rangle=\int\! d^2p\,f(\vec{p})\,|\Psi(\vec{p};t)|^2$ is the expectation value of a given function $f(\vec{p})$ of the momenta. The PCC is implicitly time-dependent, and it is normalized $|\mathcal{C}(p_1,p_2)|\leq 1$ with $+1$ ($-1$) signifying perfect correlation (anti-correlation) between the variables.
\begin{figure}[t]
    \centering
    \includegraphics[width=0.475\textwidth]{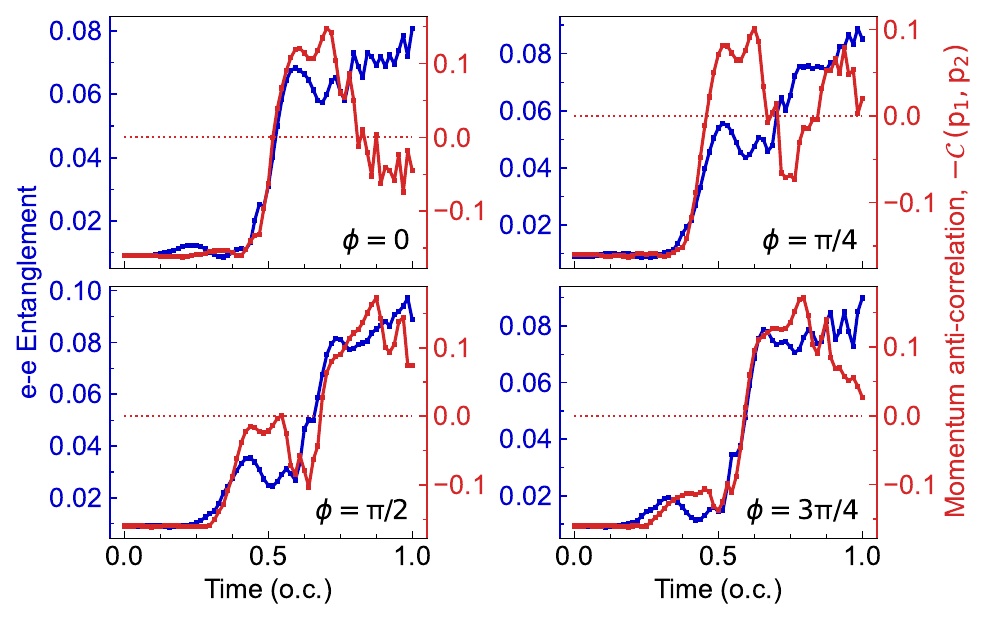}
    \caption{Time development of $e$-$e$ entanglement (blue) at four different CEP values $\phi$, as labeled, and the time-varying momentum anti-correlation function (red). The red dotted line marks $\mathcal{C}(p_1,p_2)=0$.}
    \label{fig:Schmidt-PCC-r3}
\end{figure}

Figure \ref{fig:Schmidt-PCC-r3} shows the entanglement vs.~time profile $Y(t)$ for different CEP values $\phi$, as before, but now overlaid with the momentum \emph{anti}-correlation function, $-\mathcal{C}(p_1,p_2)$. First, it is surprising how synchronized the two measures are; they rise and fall in-step and display similar sub-cycle oscillatory behavior. Initially, in the ground state of this model He atom, the electron momenta are positively correlated with $\mathcal{C}\approx 0.16$. As the interaction unfolds, we see that they rapidly become anti-correlated, which is to say that $\langle p_1 p_2\rangle < \langle p_1\rangle\langle p_2\rangle$ in the covariance. Note that the wavefunction is symmetric under exchange, and so $\langle p_1\rangle=\langle p_2\rangle$ for all time, within the tolerance for numerical error. Closer inspection of the first term in $\textrm{cov}(p_1,p_2)$ reveals that its initial value is positive: $\langle p_1 p_2\rangle\approx 4.64\times 10^{-2}\textrm{ a.u.}$, but it decreases over time by an average of $1.5\times 10^{-2}\textrm{ a.u.}$ in all cases presented in Fig.~\ref{fig:Schmidt-PCC-r3}. On the contrary, $\langle p_1\rangle\langle p_2\rangle$, which is the second, strictly-positive term in the covariance, grows over time as the electrons acquire energy from the field. It is therefore not surprising why the $\mathcal{C}(p_1,p_2)$ measure becomes negative. This decrease in $\langle p_1 p_2\rangle$ implies that the field is inducing population transfer from the aligned $(p_1 p_2>0)$ to the anti-aligned $(p_1 p_2<0)$ quadrants of momentum-space. This occurs primarily due to the RESI mechanism of nonsequential ionization, in which an $e$-$e$ rescattering event promotes the bound electron to an excited intermediate state from which it is more easily field-ionized in the subsequent half-cycle. Here, the sign reversal of $E(t)$ between the first and second half-cycles is ultimately what causes the electrons to acquire oppositely-signed momenta.

Lastly, while it is not meaningful to make direct quantitative comparisons between $Y(t)$ and $\mathcal{C}(p_1,p_2)$, we note that discrepancies between their qualitative behaviors---such as the comparatively exaggerated double-hump structure in the $\phi=\pi/4$ case of Fig.~\ref{fig:Schmidt-PCC-r3}, or a decrease in $|\mathcal{C}(p_1,p_2)|$ when the entanglement is increasing---may be due in part to the fact that the PCC is a linear correlation measure. Thus, its sensitivity to more complex relationships between the electron momenta is limited. A quantum trajectory analysis producing an ensemble of $(p_1,p_2)$ value pairs may prove beneficial in this regard, but it is beyond the scope of this contribution.
\begin{figure}[t]
    \centering
    \includegraphics[width=0.3\textwidth]{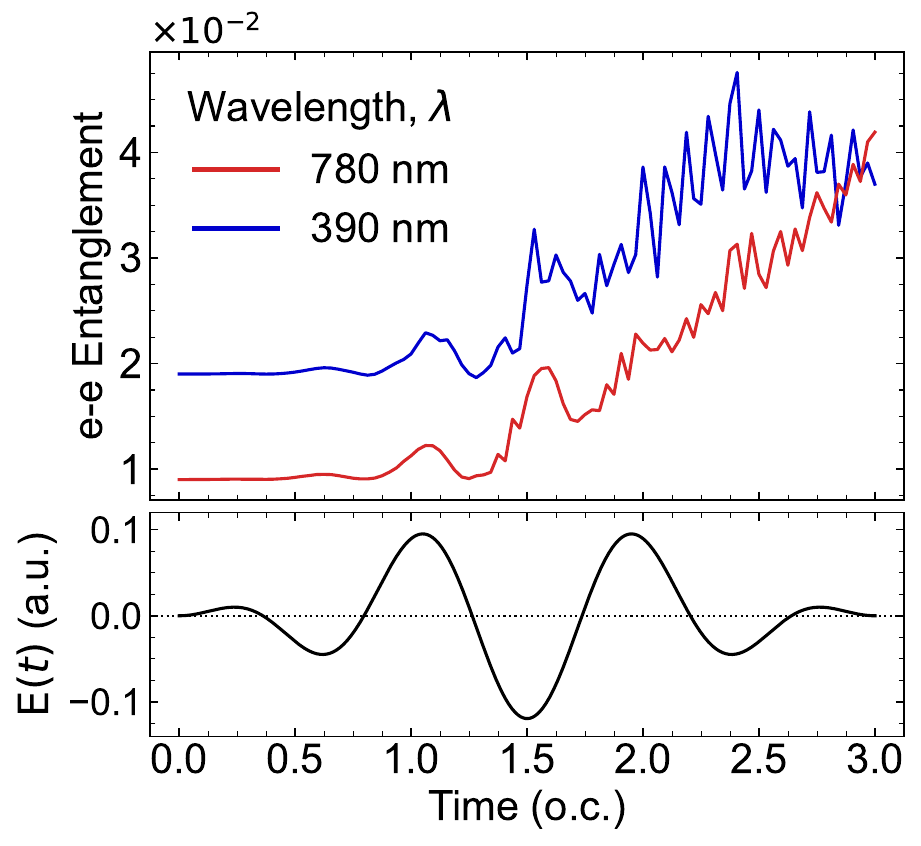}
    \caption{Top panel: Entanglement time dynamics due to a 3-cycle, $0.5\textrm{ PW/cm}^2$-intensity pulse at the indicated wavelengths. For visibility, the $\lambda=390\textrm{ nm}$ curve is offset vertically by $+1\times{10}^{-2}$. Bottom panel: The corresponding laser electric field profile.}
    \label{fig:Schmidt-lambda}
\end{figure}

\subsection{Transition to the multi-cycle regime}
In this subsection, we briefly discuss how the system behaves in the multi-cycle regime by comparing the $e$-$e$ entanglement dynamics due to a 3-cycle, $0.5\textrm{ PW/cm}^2$-intensity pulse at two different wavelengths: $\lambda=780\textrm{ nm}$ and $390\textrm{ nm}$. In Fig.~(\ref{fig:Schmidt-lambda}), observe that as the field strength begins to increase, the entanglement $Y(t)$ in both cases rises gradually and in a relatively smooth fashion as in the one-cycle case, up to about $1.5$ cycles. In the long-time limit, however, the entanglement begins to oscillate on a sub-cycle timescale while trending upwards, with the $390\textrm{ nm}$ case displaying oscillations of a larger magnitude. Following our analysis in the one-cycle regime, we conclude that this behavior originates in the wavefunction and its self-interference driven by the field, which results in spatial fringes in probability density that have varying degrees of entanglement. We also observe that the field frequency has little influence on the entanglement magnitude, at least within the beginning stages of ionization.

Last, we note that both wavelengths appear to induce entanglement oscillations whose frequencies are comparable, but this may be due in part to the sampling frequency used to calculate $Y(t)$, which is $32$ times per cycle. Performing a calculation of higher resolution, both in space and in time, may enable a relationship to be established between the frequency of the field, the spatial frequency of the probability density fringes, and the temporal frequency of the entanglement oscillations.

\section{Conclusion\label{sec:Conclusion}}
In summary, we have determined that nonsequential ionization is an important process even on short (sub-femtosecond) timescales. The time-dependent Schr\"{o}dinger equation was integrated for model atomic helium irradiated by single-cycle laser fields of varying intensities, carrier-envelope phase parameters, and frequencies. We discovered patterns between the inter-electron entanglement and momentum correlation coefficient that are interpretable based on the field profile and its ability to accelerate and subsequently induce $e$-$e$ collisions. Importantly, it was determined that in the early stages of ionization, there is an approximately constant time delay between the field attaining its maximum strength and the entanglement increasing. In addition, a connection was established between the increase in degree of entanglement, and the increase in probability current of simultaneous $2e$ escape. These new insights further elucidate the complex nature of nonsequential ionization and correlated electron emission.

\begin{acknowledgments}
The authors thank Prof.~G.~Landi for valuable discussions. This research was supported by Grant No.~DE-FG02-05ER15713 funded by the U.S.~Department of Energy, Office of Science. Calculations were performed on the BlueHive supercomputing cluster at the University of Rochester.
\end{acknowledgments}

\providecommand{\noopsort}[1]{}\providecommand{\singleletter}[1]{#1}%

\end{document}